\documentclass[twocolumn, showpacs, aps, prd]{revtex4-1}
\usepackage{mathrsfs}
\usepackage{amssymb}
\usepackage{latexsym}
\usepackage{hyperref}
\usepackage{amsmath}
\usepackage{graphicx}

\begin{document}
\title{Self-consistency in relativistic theory of infinite statistics fields}
\author{Chao Cao}\email{ccldyq@gmail.com}
\author{Yi-Xin Chen}\email{yxchen@zimp.zju.edu.cn}
\author{Jian-Long Li}\email{marryrene@gmail.com}
\affiliation{Zhejiang Institute of Modern Physics, Zhejiang
University, Hangzhou 310027, China}

\begin{abstract}
Infinite statistics in which all representations of the symmetric
group can occur is known as a special case of quon theory. Our
previous work has built a relativistic quantum field theory which
allows interactions involving infinite statistics particles. In this
paper, a more detailed analysis of this theory is available. Topics
discussed include cluster decomposition, CPT symmetry and
renormalization.
\end{abstract}
 \maketitle
\section{Introduction}
Most conventional quantum theories are based on Bose-Einstein
statistics and/or Fermi-Dirac statistics, i.e. only two
1-dimensional representations of the permutation group are allowed.
However the general principles of quantum theory do not have this
requirement \cite{Messiah}, and Bose (Fermi) statistics can be
violated by a small amount. One famous approach of such violation is
quon theory, in which the basic algebra can be obtained as the
convex sum of Bose and Fermi statistics \cite{Greenberg:1991ec}.
Especially when the sum is the average of these two algebras, we can
get a special statistics called infinite statistics
\cite{Greenberg:1989ty}.

The basic algebra of infinite statistics is
$a_ka^{\dag}_l=\delta_{kl}$, which involves no commutation relation
between annihilation and creation operators. The quantum states are
orthogonal under any permutation of the identical particles. So it
allows all representations of the symmetric group to occur.
Furthermore, the loss of local commutativity also implies violation
of locality, which is an important characteristic of quantum
gravity. By virtue of these properties, infinite statistics has been
applied to many subjects, such as black hole statistics
 \cite{Strominger:1993si,Volovich:1996gw,Minic:1997ym}, dark energy
quanta
 \cite{Ng:2007bp,Ng:2008pi,Jejjala:2007hh,Li:2008qh,Medved:2008vh},
large N matrix theory
 \cite{Douglas:1994kw,Gopakumar:1994iq,Aref'eva:1995pz} and
holography principle \cite{Shevchenko:2008zy,Chen:2008rz}. Many of
these applications involve discussions in relativistic case.

However there are some difficulties for infinite statistics to have
a consistent relativistic theory. Due to the lack of local
commutativity, Lorentz invariance of the $S$-matrix is unapparent in
the Dyson formula \cite{Greenberg:1991ec,Greenberg:1989ty}. And the
conservation of statistics also acquires some special form the the
interaction Hamiltonian \cite{Chow:2000cm}. Our previous work
\cite{Cao:2009mi} constructs a relativistic quantum field theory
obeying infinite statistics by solving the two difficulties above.

In the present work we keep on investigating the self-consistency of
this theory. Greenberg has showed that cluster decomposition and the
CPT theorem hold for free fields \cite{Greenberg:1991ec}. In this
paper, we show that the clustering still holds for interacting
theory, but CPT symmetry violates in vector-spinor interactions. We
also discuss the renormalizability of this theory by using
non-perturbative methods.

This paper is organized as follows. In Sec. \ref{s2} we introduce
the elementary ingredients of infinite statistics and main results
of our previous work. In Sec. \ref{s3} we prove the cluster
decomposition for our new theory. In Sec. \ref{s4} we discuss the
CPT symmetry. In Sec. \ref{s5} we discuss the renormalizability of
infinite statistics theory. General conclusions are given in Sec.
\ref{s6}. We also introduce a non-perturbative method for infinite
statistics field theory in Appendix.
\section{Infinite statistics}\label{s2}
The quon algebra is the convex sum of Bose and Fermi statistics
\begin{equation}\label{1}
\frac{1+q}{2}[a_k,a^{\dag}_l]+\frac{1-q}{2}[a_k,a^{\dag}_l]_+=\delta_{kl},
\end{equation}
convexity requires $0\leq q\leq 1$. When $q=+1$ ($q=-1$), this
becomes Bose (Fermi) statistics. The quon statistics interpolates
smoothly between Bose and Fermi statistics when $q$ transfers from
$+1$ to $-1$. Especially when $q=0$ each statistics has equal weight
and all  representations of the symmetric group can occur. We call
it infinite statistics, in which the basic algebra is
\begin{equation}\label{2}
a_ka^{\dag}_l=\delta_{kl}.
\end{equation}
We can get a Fock-state representation by defining a unique vacuum
state annihilated by all the annihilators
\begin{equation}\label{3}
a_k|0\rangle=0.
\end{equation}
The m-particle state is constructed as
\begin{equation} \label{4}
|\phi_m\rangle=(a_{k_1}^{\dag})^{m_1}(a_{k_2}^{\dag})^{m_2}...(a_{k_j}^{\dag})^{m_j}|0\rangle
\end{equation}
with $m_1+m_2+...+m_j=m$, where we have $a^{\dagger}_{k_i} \ne
a^{\dagger}_{k_{i+1}}$. Such states have positive norms and the
normalization factor equals one. Since there is no commutation
relation between two annihilation or creation operators, the states
created by any permutations of creation operators are orthogonal.
That's why it is also called ¡°quantum Boltzmann statistics¡±.

One can define a set of number operators $\hat{n}_i$ such that
\begin{equation} \label{5}
[\hat{n}_i, a_j]=-\delta_{ij}a_j.
\end{equation}
Then the total number operator is $N=\sum\limits_{i}\hat{n}_i$, and
the energy operator is given by
$E=\sum\limits_{i}\epsilon_i\hat{n}_i$, where $\epsilon_i$ is the
single particle energy. The explicit form of $\hat{n}_i$ is
\begin{equation} \label{6}
\begin{split}
\hat{n}_i=a_i^{\dag}a_i+\sum\limits_{k}a_k^{\dag}a_i^{\dag}a_ia_k+\sum\limits_{k_1,k_2}a_{k_1}^{\dag}a_{k_2}^{\dag}a_i^{\dag}a_ia_{k_2}a_{k_1}+\cdots\\
+\sum\limits_{k_1,k_2,\ldots,k_s}a_{k_1}^{\dag}a_{k_2}^{\dag}\cdots
a_{k_s}^{\dag}a_i^{\dag}a_ia_{k_s}\cdots a_{k_2}a_{k_1}+\cdots,
\end{split}
\end{equation}
which is obviously a non-local operator. One can directly use
relation (\ref{2}) to check that this definition obeys Eq.
({\ref{5}}).

All the above discussion is under non-relativistic case. It's not
difficult to extend these to the relativistic field theory. We can
construct infinite statistics field that transform irreducibly under
the Lorentz group in the same way of conventional Bose (Fermi)
fields construction. By using the general transformation rules of
particle states and the Fourier transform, the annihilation field
$\psi^+_l(x)$ and creation field $\psi^-_l(x)$ with mass $m$ in
momentum space are
\begin{equation} \label{7}
\psi^{+(n)}_l(x)=\sum\limits_{\sigma n}(2\pi)^{-3/2}\int d^3p\
u_l^{(n)}({\boldsymbol p},\sigma)e^{ip\cdot x}a_{\boldsymbol
p}^{(n)}(\sigma), \end{equation}
\begin{equation}\label{8}
\psi^{-(n)}_l(x)=\sum\limits_{\sigma n}(2\pi)^{-3/2}\int d^3p\
v_l^{(n)}({\boldsymbol p},\sigma)e^{-ip\cdot
x}a^{\dag(n)}_{\boldsymbol p}(\sigma),
\end{equation}
Here we have chosen relativistic four-vector notation
$p^\mu=(p^0,{\boldsymbol p})$ ($p^0\equiv\sqrt{{\boldsymbol
p}^2+m^2}$), $\sigma$ labels spin z-components (or helicity for
massless particles), and the superscript $(n)$ labels particle
species
\begin{equation}\label{9}
a_{\boldsymbol p}^{(n)}(\sigma)a_{\boldsymbol
p'}^{\dag(n')}(\sigma')=\delta(nn')\delta(\sigma\sigma')\delta^3({\boldsymbol
p}-{\boldsymbol p'}).
\end{equation}
In a theory based on infinite
statistics the local commutativity ($[\psi(x), \psi^{\dag}(y)]_{\mp}
=0$ for $x-y$ spacelike) does not hold, so we can't uniquely
determine a linear combination field
$\psi(x)=\kappa\psi^+(x)+\lambda\psi^{-c}(x)$, so the basic field in
this theory should be $\psi^+(x)$ and $\psi^-(x)$.

For a relativistic field theory, the $S$-matrix should be Lorentz
invariant,
\begin{equation}\label{10}
\begin{aligned}
S&=T\{{\rm exp}(-i\int_{-\infty}^{\infty}dtV(t))\}\\
 &=1+\sum\limits_{N=1}^{\infty}\frac{(-i)^N}{N!}\int d^4x_1\cdots
 d^4x_N T\{\mathscr{H}(x_1)\cdots \mathscr{H}(x_N)\},
\end{aligned}
\end{equation}
In which $V(t)=\int d^3x \mathscr{H}(x)$ is the interaction term
$H=H_0+V$ and $T\{\ \}$ denotes the time-ordered product. It's not
difficult to construct scalar interaction density $\mathscr{H}(x)$
out of creation and annihilation operators (\ref{7}) and (\ref{8}).
The problem arises from the time-ordering of the operator product. A
direct way is to require $\mathscr{H}(x)$ all commute at spacelike
separations
\begin{equation}\label{11}
[\mathscr{H}(x),\mathscr{H}(x')]=0\ \ {\rm for}\ \ (x-x')^2\geq 0.
\end{equation}
However this condition is to strong for infinite statistics fields.
In our previous work \cite{Cao:2009mi} we proved that the relation
(\ref{9}) leads to a weaker locality condition
\begin{equation}\label{12}
\int d^3y \ \boldsymbol{x}\ [\mathscr{H}({\boldsymbol
x},0),\mathscr{H} ({\boldsymbol y},0)]=0.
\end{equation}
Since the weakest sufficient condition for Lorentz invariance of the
$S$-matrix is given in \cite{weinberg} that $0=\int d^3x\int d^3y\
{\boldsymbol x}\ [\mathscr{H}({\boldsymbol
x},0),\mathscr{H}({\boldsymbol y},0)]$, we conclude that the
interaction field theory based on infinite statistics is Lorentz
invariant. It's worth noticing that this invariance is only valid
for whole structure of $S$-matrix, and the perturbative terms in
Dyson series (\ref{10}) (from second order to higher) do not hold
such invariance. This is not strange to us, because infinite
statistics theory is non-local, and we can not limit the number of
interaction points, so each order of $S$-matrix can not solely
describe a real physical process. However, we can analyze the
perturbative graphs as an analogy tool, as we will see in
Sec.{\ref{s5}}.

Another problem for relativistic theory of infinite statistics
fields comes from the condition that the energy of widely spacelike
separated subsystems should be additive
\begin{equation}\label{13}
[\mathscr{H}(x), \psi(x')]\rightarrow 0, \ {\rm as} \
x-x'\rightarrow \infty \ {\rm spacelike}.
\end{equation}
We figured out this problem by introducing a general operator
definition $\mathcal{O}\rightarrow \mathcal{A}(\mathcal{O})$ (this
new definition also satisfies the weak locality condition
(\ref{12})) in infinite statistics theory
\begin{equation}\label{14}
\begin{aligned}
\mathcal {A}(\mathcal
{O})\equiv&\sum\limits_{m=0}^{\infty}\sum\limits_{n_1,...,n_m,}\sum\limits_{\sigma_1,\cdots,\sigma_m}\int
d^3{k_1}\cdots d^3{k_m}\\&a^{\dag(n_1)}_{{\boldsymbol
k}_1}(\sigma_1)\cdots a^{\dag(n_m)}_{{\boldsymbol
k}_m}(\sigma_m)\mathcal{O}\ a^{(n_m)}_{{\boldsymbol
k}_m}(\sigma_m)\cdots a^{(n_1)}_{{\boldsymbol k}_1}(\sigma_1)
\end{aligned}
\end{equation}
and requiring that $\mathscr{H}$ in interaction density
$\mathcal{A}(\mathscr{H}(x))$ should have at least one annihilation
infinite statistics field and one creation infinite statistics
field. Then we can get
\begin{equation}\label{15}
[\mathcal{A}(\mathscr{H}(x)), \psi(x')]\rightarrow 0, \ {\rm as} \
x-x'\rightarrow \infty \ {\rm spacelike}.
\end{equation}
We also showed this condition imposes conservation of statistics
rules, i.e. there must be infinite statistics particles both in the
initial and final states when an interaction involves infinite
statistics particles.
\section{cluster decomposition}\label{s3}
Cluster decomposition is one of the fundamental principles of
physics, which indicates that widely spacelike separated experiments
have unrelated results. This is the foundation that we can make
predictions about experiments. Greenberg
\cite{Greenberg:1991ec,Greenberg:1989ty} has showed that an
arbitrary vacuum matrix element of a product of infinite statistics
fields is a sum of products of two-point functions, which means
cluster decomposition principle holds for free propagating fields.

In general scattering theory, this principle states that for distant
processes $\alpha_1\rightarrow\beta_1$,
$\alpha_2\rightarrow\beta_2$, $\cdots$,
$\alpha_n\rightarrow\beta_n$, the overall $S$-matrix element can be
factorized
\begin{equation}\label{16}
S_{\beta_1\beta_2\cdots\beta_n,\alpha_1\alpha_2\cdots\alpha_n}\rightarrow
S_{\beta_1\alpha_1}S_{\beta_2\alpha_2}\cdots S_{\beta_n\alpha_n},
\end{equation}
or more generally
\begin{equation}\label{17}
S_{\beta\alpha}=\sum\limits_{\rm PART}(\pm)
S^C_{\beta_1\alpha_1}S^C_{\beta_2\alpha_2}\cdots.
\end{equation}
The sum is over all clusters $\alpha_1$, $\alpha_2$, $\cdots$
(likewise $\beta_1$, $\beta_2$, $\cdots$)in the state $\alpha$
(likewise $\beta$). The term $S^C_{\beta\alpha}$ with superscript
$C$ is the connected part of the $S$-matrix. In conventional theory
the minus sign denotes permutations of odd fermions, however as we
will see in the next section, due to CPT symmetry, such sign also
involves in infinite statistics spinor interactions.

In order to prove clustering, we can take two steps. First,
factorize the particles in the state $\alpha$, $\beta$ into
different products. Then prove that every term including spatially
distant processes vanishes. It's not difficult to realize these in
infinite statistics field theory. As we showed in \cite{Cao:2009mi},
due to the non-local operator form $\mathcal{A}(\mathscr{H})$, the
fields in $\mathscr{H}$ can always be transported by the creation or
annihilation operators $a$, $a^{\dag}$ in $\int d^3p_i \cdots
a^{\dag}_{{\boldsymbol p}_i}\cdots\mathscr{H}(x)\cdots
a_{{\boldsymbol p}_i}\cdots$ (the creation fields move to the left,
and the annihilation fields move to the right). On the other hand,
the conservation of statistics rules require that each $\mathscr{H}$
has at least one creation and annihilation fields. So one
$\mathscr{H}(x)$ in interaction Hamiltonian density
$\mathcal{A}(\mathscr{H}(x))$ at coordinate $x$ can always interact
with another $\mathscr{H}(x')$ at coordinate $x'$. This is our first
step. We have also showed that in infinite statistics
$\psi^{+(n)}(x)\psi^{-(m)}(x')\sim
\delta(nm)\Delta_+(x-x')\rightarrow 0\ \  {\rm as} \ x-x'\rightarrow
\infty \ {\rm spacelike}$. Since the connection of two
$\mathscr{H}$s comes from the contraction between creation and
annihilation fields, we can see that the correlations among very
distant experiments vanish. So the cluster decomposition principle
is also valid in our new theory.
\section{CPT symmetry}\label{s4}
As we have seen in Sec. \ref{s2}, the construction of free infinite
statistics fields is in the same way as we do in conventional field
theory. The only difference is the basic fields here are just
creation and annihilation fields. So the CPT transformations of
infinite statistics fields are similar to boson (fermion) fields.
For a scalar, vector, or spinor field the transformation rule is
\cite{weinberg}
\begin{eqnarray}
&[\rm {CPT}]\phi^{\pm}(x)[\rm {CPT}]^{-1}=\phi^{\pm
c}(-x)\label{18},\\
&[\rm {CPT}]\phi^{\pm\mu}(x)[\rm {CPT}]^{-1}=-\phi^{\pm\mu
c}(-x)\label{19},\\
&[\rm {CPT}]\psi^{\pm}(x)[\rm {CPT}]^{-1}=-\gamma_5\psi^{\mp
c*}(-x)\label{20},
\end{eqnarray}
in which
$\gamma_5\equiv-i\gamma^0\gamma^1\gamma^2\gamma^3=\begin{bmatrix}1
&0\\0 &-1\end{bmatrix},\gamma^0=-i\begin{bmatrix}0 &1\\1
&0\end{bmatrix},{\boldsymbol \gamma}=-i\begin{bmatrix}0&{\boldsymbol
\sigma}\\-{\boldsymbol \sigma} &0\end{bmatrix}$, the components of
$\boldsymbol{\sigma}$ are the usual Pauli matrices. Since a scalar
interaction density $\mathcal{A}(\mathscr{H}(x))$ must be formed
from tensors with an even total number of spacetime indices, the
minus sign in Eq. (\ref{19}) has no effect on CPT symmetry of
$\mathcal{A}(\mathscr{H}(x))$. For scalar and vector fields if every
term in $\mathscr{H}$ has a corresponding antiparticle part (e.g. if
there is one term $\phi^-\phi^{+c}\phi^{+}$, then there must also
exists another term $\phi^{-c}\phi^{+}\phi^{+c}$), we have
\begin{equation}\label{21}
[\rm{CPT}]\mathcal{A}(\mathscr{H}(x))[\rm{CPT}]^{-1}=\mathcal{A}(\mathscr{H}(-x)).
\end{equation}
From Eq. (\ref{10}) we can see that CPT commutes with the
$S$-matrix, i.e. the CPT theorem holds for scalar and vector fields
if we take a natural limit (the particle and antiparticle terms
appear symmetrically) on $\mathscr{H}(x)$ in interaction density
$\mathcal{A}(\mathscr{H}(x))$.

The case is a bit more complicated for spinor fields. In order to
construct scalar interaction densities out of spinor fields, we have
to use the bilinear combination $\bar{\psi}_1(x)M\psi_2(x)$, in
which $\bar{\psi}^{\pm}\equiv \psi^{\pm {\dag}}\beta,\beta\equiv
i\gamma^0$, the subscripts $1,2$ include indices such as
annihilation/creation index $\pm$ and particle/antiparticle index
$c$. When the matrix $M$ takes $\boldsymbol 1$, $\gamma^{\mu}$,
$\mathscr{J}^{\mu\nu}\equiv-\frac{1}{4}[\gamma^{\mu},\gamma^{\nu}]$,
$\gamma_5\gamma^{\mu}$, or $\gamma_5$, such bilinear transfers as a
scalar, vector, tensor, axial vector, and pseudoscalar respectively.
By applying Eq. (\ref{21}) we can get
\begin{equation}\label{22}
\begin{aligned}
&\left[{\rm CPT}\right][\bar{\psi}_1(x)M\psi_2(x)][{\rm
CPT}]^{-1}\\=&\psi_1^{(-)(c)T}(-x)\gamma_5\beta
M^*\gamma_5\psi_2^{(-)(c)*}(-x)\\=&-\psi_1^{(-)(c)T}(-x)\beta(\gamma_5
M^*\gamma_5)\psi_2^{(-)(c)*}(-x),
\end{aligned}
\end{equation}
in which the superscript $(-)$ denotes the interchange between
creation field and annihilation field, the superscript $(c)$ denotes
the interchange between particle and antiparticle. The complex
conjugate of $M$ in the right hand side of Eq. (\ref{22}) comes from
the antiunitary of CPT. It's worth noticing that $\gamma_5 M
\gamma_5=(-1)^nM$, where $n$ is the tensor rank of
$\bar{\psi}_1(x)M\psi_2(x)$, since the interaction density should be
scalar, the sign $(-1)^n$ does not affect CPT symmetry. Especially,
when $M$ takes unit matrix $\boldsymbol 1$ or $\gamma_5$, we have
$M^*=M$, then by using Eq. (\ref{22}) and considering the
hermiticity of interaction density, the general form of
$\mathscr{H}$ to keep CPT invariance is
\begin{equation}\label{23}
\begin{aligned}
&C_1(\psi^{+\dag}\beta M\psi^+-\psi^{-cT}\beta
M\psi^{-c*})\\+&C_2(\psi^{+\dag}\beta M\psi^{-c}-\psi^{-cT}\beta
M\psi^{+*})\\+&C_3(\psi^{-c\dag}\beta M\psi^{+}-\psi^{+T}\beta
M\psi^{-c*})\\+&C_4(\psi^{+\dag}\beta M\psi^{+*}-\psi^{-c T}\beta
M\psi^{-c})\\+&C_5(\psi^{+T}\beta M\psi^+-\psi^{-c*T}\beta
M\psi^{-c*}),
\end{aligned}
\end{equation}
in which $C_1,\cdots,C_5$ are coefficients, the minus sign appears
in such interactions, just like conventional fermion field theories.

When $M$ takes $\gamma^{\mu}$, $\mathscr{J}^{\mu\nu}$ or
$\gamma_5\gamma^{\mu}$,  we have $M^*\neq M$ and
$\bar{\psi}_1(x)M^*\psi_2(x)$ does not satisfy the Lorentz
transformation rules. So the Lorentz invariance indicates CPT
violation in such kinds of interactions. This is different with
conventional Dirac fields, in which the anticommutation of fermionic
operators gives
\begin{equation}\label{24}
\psi_1^T(-x)\gamma_5\beta
M^*\gamma_5\psi^*_2(-x)=[\bar{\psi}_1(-x)\gamma_5
M\gamma_5\psi_2(-x)]^{\dag},
\end{equation}
then the hermiticity ensures CPT invariance. However Eq. (\ref{24})
does not hold for infinite statistics fields without commutation
relation. Noting that term $\bar{\psi}_1(x)\gamma^{\mu}\psi_2(x)$ is
necessary in vector-spinor couplings, we conclude the
``electromagnetic" interactions for infinite statistics field theory
have violated CPT symmetry.

\section{renormalization}\label{s5}
In quantum field theory, the normalizability is usually related to
the superficial degree of divergence $D$, which is the actual degree
of divergence of the integration over the region of momentum space
in which the momenta of all internal lines go to infinity together.
By considering the structure of Feynman diagrams (or dimensional
analysis) \cite{weinberg}, we can get
\begin{equation}\label{25}
D=4-\sum\limits_{f}E_{f}(s_f+1)-\sum\limits_{i}N_i\Delta_i,
\end{equation}
where $E_f$ denotes the number of external lines of field type $f$,
$s_{f}$ denotes the spin of field type $f$, $N_i$ denotes the number
of vertices of interaction type $i$. $\Delta_i$ is a parameter
characterizing interactions of type $i$ (i.e. the dimensionality of
coupling coefficient), $\Delta_i\equiv 4-d_i-\sum\limits_f
n_{if}(s_f+1)$, in which $d_i$ denotes the number of derivatives in
each interaction of type $i$, $n_{if}$ denotes the number of fields
of type $f$ in interactions of type $i$.

Since the term $4-\sum\limits_{f}E_{f}(s_f+1)$ is fixed for a
specific physical process, the divergence is mainly depended on
$\Delta_i$. If $\Delta_i\geq 0$, then $D$ has a upper bound $D\leq
4-\sum\limits_{f}E_f(s_f+1)$. In this case only a finite number of
divergences can exists, we can removed such divergences by a
renormalization of fields, so we call such theories renormalizable.
If $\Delta_i<0$, the superficial degree of divergence will become
larger when more vertices involves, we need an infinite number of
couplings to absorb such divergence. Such theories are called
non-renormalizable.

As we have shown in \cite{Cao:2009mi}, the external line term in
infinite statistics theory has the same form as in conventional
theory, while the internal line term has a non-covariant form
\cite{footnote}
\begin{equation}\label{26}
\begin{aligned}
\Delta_{F}(q)&\equiv(2\pi)^{-4}\frac{-P_{lm}^{(L)}(q)}{2\sqrt{{\boldsymbol
q}^2+m^2}(q^0-\sqrt{{\boldsymbol
q}^2+m^2}+i\epsilon)}\\&=(2\pi)^{-4}(\frac{1}{2}+\frac{q^0}{2\sqrt{{\boldsymbol
q}^2+m^2}})\frac{P_{lm}^{(L)}(q)}{q^2+m^2-i\epsilon}.
\end{aligned}
\end{equation}
Noting that the propagator has the same dimensionality as
conventional theory, especially it has the same behavior near its
pole as for a boson/fermion field, so we can use the traditional
dimensional analysis method to calculate $D$ and $\Delta_i$ for
infinite statistics field theory.

We can deal with the finite number of divergences by
non-perturbative methods (see Appendix for details). For example,
consider a vector particle self-energy process, with two vector
external lines and without any spinor external lines. It has $D=2$,
the divergent part is a second-order polynomial in $q$. The complete
propagator is given by a sum of one-particle-irreducible (OPI)
subgraphs,
\begin{equation}\label{27}
\begin{aligned}
\Delta'_{\mu\nu}(q)&=\Delta_{\mu\nu}(q)+\Delta_{\mu\rho}(q)\Pi^{\rho\sigma}(q)\Delta_{\sigma\nu}(q)\\&\
+\Delta_{\mu\rho}(q)\Pi^{\rho\sigma}(q)\Delta_{\sigma\alpha}(q)\Pi^{\alpha\beta}(q)\Delta_{\beta\nu}(q)+\cdots\\&=[\Delta(q)^{-1}-\Pi(q)]^{-1}_{\mu\nu},
\end{aligned}
\end{equation}
in which
$\Delta_{\mu\nu}(q)=(\frac{1}{2}+\frac{q^0}{2\sqrt{{\boldsymbol
q}^2}})\frac{(P^{(L)}_{lm}(q))_{\mu\nu}}{q^2-i\epsilon}$ is the bare
vector field propagator, $\Pi^{\mu\nu}(q)$ is the OPI contribution.
As we have proved, $\Delta'_{\mu\nu}(q)$ is Lorentz covariant, while
$\Delta_{\mu\nu}(q)$ and $\Pi^{\mu\nu}(q)$ is non-covariant.
However, we can do a transformation
\begin{equation}\label{28}
\Delta^*_{\mu\nu}(q)\equiv
\frac{(P^{(L)}_{lm}(q))_{\mu\nu}}{q^2-i\epsilon},
\end{equation}
\begin{equation}\label{29}
\begin{aligned}
\Pi^{*\mu\nu}(q)&\equiv
\Pi^{\mu\nu}(q)+{\Delta^*_{\mu\nu}(q)}^{-1}-{\Delta_{\mu\nu}(q)}^{-1}\\&=\Pi^{\mu\nu}(q)-(\sqrt{{\boldsymbol
q}^2}-{q^0})^2[(P^{(L)}_{lm}(q))_{\mu\nu}]^{-1},
\end{aligned}
\end{equation}
then
\begin{equation}\label{30}
\Delta'_{\mu\nu}(q)=[\Delta(q)^{-1}-\Pi(q)]^{-1}_{\mu\nu}=[{\Delta^*(q)}^{-1}-\Pi^*(q)]^{-1}_{\mu\nu}.
\end{equation}
Now both ${\Delta_{\mu\nu}^*(q)}$ and $\Pi^{*\mu\nu}(q)$ are Lorentz
covariant, and we can use this property and $q_{\mu}\Pi^{*\mu\nu}=0$
to write the form of $\Pi(q)$
\begin{equation}\label{31}
\begin{aligned}
\Pi^{\mu\nu}(q)=&(\eta^{\mu\nu}q^2-q^{\mu}q^{\nu})(\pi(q^2))\\&-(\sqrt{{\boldsymbol
q}^2}-{q^0})^2[({P^{(L)}_{lm}(q)})^{-1}]^{\mu\nu},
\end{aligned}
\end{equation}
The non-covariant term is finite and become zero on the mass-shell.
Since the introducing of OPI should not change the structure of the
pole at $q^2=0$, so $\pi(0)=0$. Now we can renormalize the vector
field $A^{\mu}\rightarrow Z^{-1/2}A^{\mu}$, then
$\pi(q^2)=1-Z+\pi_{\rm LOOP}(q^2)$ and we can get $Z=1+\pi_{\rm
LOOP}(0)$.

We see that the divergence in such vector-spinor interaction graph
is absorbed by coupling constant $Z$. One can easily check that the
self-energy process in scalar electrodynamics (in which CPT theorem
holds) can also be renormalized in a similar way.

\section{conclusions\label{s6}}
Quantum field theory based on infinite statistics is a valid
relativistic theory. In this paper, we analyze the self-consistency
of such theory. Although this is a non-local theory, it obeys
cluster decomposition principle. If we take a limit that the
particle and antiparticle terms appear symmetrically in the
interaction density, the CPT theorem also holds when we realize
infinite statistics theory by scalar fields, vector fields or scalar
(pseudoscalar) spinor-pairs $\bar{\psi}_1\psi_2$
($\bar{\psi}_1\gamma_5\psi_2$). However, Lorentz invariance
indicates CPT violation in interactions involving vector (axial
vector, tensor) spinor-pairs $\bar{\psi}_1\gamma^{\mu}\psi_2$
($\bar{\psi}_1\gamma_5\gamma^{\mu}\psi_2$,
$\bar{\psi}_1\mathscr{J}^{\mu\nu}\psi_2$), which means that the
``electromagnetic" interactions have violated CPT symmetry. We also
showed that this theory is renormalizable through dimension analysis
and non-perturbative methods.

\acknowledgments

This work is supported in part by the NNSF of China Grant No.
90503009, No. 10775116, and 973 Program Grant No. 2005CB724508.

\appendix*\section{Non-Perturbative Methods}
In this Appendix, we give a non-perturbative method which will be
useful in deriving results valid to whole $S$-matrix (beyond
perturbation theory). The analysis is analog to Chap. 10 in
\cite{weinberg}, however one should keep in mind the difference
between infinite statistics theory and conventional field theory.

Consider a momentum-space amplitude
\begin{equation}\label{A1}
\begin{aligned}
G(q_1\cdots q_n)\equiv&\int d^4x_1\cdots d^4x_ne^{-iq_1\cdot
x_1}\cdots e^{-iq_n\cdot x_n}\\&\langle T\{A_1(x_1)\cdots
A_n(x_n)\}\rangle_0,
\end{aligned}
\end{equation}
where $A$s are Heisenberg-picture operators and
$\langle\cdots\rangle_0$ denotes the expectation value in the true
vacuum. By using Fourier representation of the step function
$\theta(\tau)=-\frac{1}{2\pi
i}\int^{\infty}_{\infty}\frac{d\omega\exp{-i\omega
\tau}}{\omega+i\epsilon}$ and spacetime translational invariance, we
can show that near the pole (see Chap. 10 in \cite{weinberg} for
details)
\begin{equation}\label{A2}
\begin{aligned}
G\rightarrow&\frac{-2i\sqrt{\boldsymbol{q}^2+m^2}}{q^2+m^2-i\epsilon}(2\pi)^7\delta^4(q_1+\cdots+q_n)
\\&\sum\limits_{\sigma}M_{0|\boldsymbol{q},\sigma}(q_2\cdots
q_r)M_{\boldsymbol{q},\sigma|0}(q_{r+2}\cdots q_n)+\rm{OT},
\end{aligned}
\end{equation}
where
\begin{equation}\label{A3}
q\equiv q_1+\cdots+q_r=-q_{r+1}-\cdots-q_n,\ 1\leq r\leq n-1,
\end{equation}
\begin{equation}\label{A4}
\begin{aligned}
\ &\int d^4x_1\cdots d^4x_r e^{-iq_1\cdot x_1}\cdots e^{-iq_r\cdot
x_r}\\&\times(\Psi_0, T\{A_1(x_1)\cdots
A_r(x_r)\}\Psi_{\boldsymbol{p},\sigma})\\=&(2\pi)^4\delta^4(q_1+\cdots+q_r-p)M_{0|\boldsymbol{p},\sigma}(q_2\cdots
q_r),
\\\ &\int d^4x_{r+1}\cdots d^4x_n e^{-iq_{r+1}\cdot
x_{r+1}}\cdots e^{-iq_n\cdot
x_n}\\&\times(\Psi_{\boldsymbol{p},\sigma},
T\{A_{r+1}(x_{r+1})\cdots
A_n(x_n)\}\Psi_0)\\=&(2\pi)^4\delta^4(q_{r+1}+\cdots+q_n-p)M_{\boldsymbol{p},\sigma|0}(q_{r+2}\cdots
q_n).
\end{aligned}
\end{equation}
`OT' denotes other terms that exhibit different poles. This
calculation is irrelative to the kind of particle statistics, and
the pole structure of $G$ is quite like the pole in a Feynman
diagram with a single internal line. In conventional theory,
$M_{0|\boldsymbol{p},\sigma}(q_2\cdots q_r)$ is explained as $r$
external lines (with the factor $(2\pi)^{-2/3}u_l$ stripped away).
Since $A(x)$ need not be some elementary particle field, it may be
also a bound state, so the pole arises not from single Feynman
diagrams and $G$ can be seen as a contribution of infinite sums of
diagram. While in infinite statistics theory, the time-ordered
product in $M$ is usually not Lorentz covariant, so we can not
directly explain it as the external line term. However we can
further reduce Eq. (\ref{A2}) to
\begin{equation}\label{A5}
\begin{aligned}
&G\rightarrow \delta^4(q_1+\cdots+q_n)
\\&\times\sum\limits_{\sigma_1,\cdots,\sigma_{n-1}}(\Psi_0, A_1(0)\Psi_{\boldsymbol{q}_1,\sigma_1})(\Psi_{\boldsymbol{q}_1,\sigma_1},A_2(0)\Psi_{\boldsymbol{q}_1+\boldsymbol{q}_2,\sigma_1})\cdots\\&\times(\Psi_{\boldsymbol{q}_1+\cdots+\boldsymbol{q}_{n-1},\sigma_{n-1}},A_n(0)\Psi_{0})
\\&\times\frac{-2i\sqrt{\boldsymbol{q_1}^2+m_1^2}}{q_1^2+m_1^2-i\epsilon}\cdot\frac{-2i\sqrt{(\boldsymbol{q}_1+\boldsymbol{q}_2)^2+m_2^2}}{(q_1+q_2)^2+m_2^2-i\epsilon}\cdots\\&\times\frac{-2i\sqrt{(\boldsymbol{q}_1+\cdots+\boldsymbol{q}_{n-1})^2+m_{n-1}^2}}{(q_1+\cdots
q_{n-1})^2+m_{n-1}^2-i\epsilon}+\rm{OT},
\end{aligned}
\end{equation}
in which every matrix element contains only one operator and should
be Lorentz covariant. So Eq. (\ref{A5}) can be seen as a tree level
non-perturbative diagram with $n$ external lines.

First we see the $n=2$ case. Eq. (\ref{A5}) becomes
\begin{equation}\label{A6}
\begin{aligned}
&\ \ \int d^4 x_1 d^4 x_2 e^{-iq_1\cdot x_1}e^{-i q_2\cdot
x_2}(\Psi_0,T\{\mathcal{O}_l(x_1)\mathcal{O}^{\dag}_{l'}(x_2)\}\Psi_0)\\&\xrightarrow{q_1^0=\sqrt{\boldsymbol{q}_1^2+m^2}}
(2\pi)^7\frac{-2i\sqrt{\boldsymbol{q}_1^2+m^2}}{q_1^2+m^2-i\epsilon}\\\times&\sum\limits_{\sigma}(\Psi_0,\mathcal{O}_l(0)\Psi_{\boldsymbol{q}_1,\sigma})(\Psi_{\boldsymbol{q}_1,\sigma},\mathcal{O}^{\dag}_{l'}(0)\Psi_0)\delta^4(q_1+q_2)
\\=&\frac{-2i|N|^2\sqrt{\boldsymbol{q}_1^2+m^2}}{q_1^2+m^2-i\epsilon}\sum\limits_{\sigma}u_l(\boldsymbol{q}_1,\sigma)u^*_{l'}(\boldsymbol{q}_1,\sigma)(2\pi)^4\delta^4(q_1+q_2),
\end{aligned}
\end{equation}
in which $\mathcal{O}(x)$ is a Heisenberg-picture operator with the
same Lorentz transformation properties as some sort of free field
$\psi_l$, so we can write matrix element
\begin{equation}\label{A7}
(\Psi_0,\mathcal{O}_l(0)\Psi_{\boldsymbol{q},\sigma})=(2\pi)^{-3/2}Nu_l(\boldsymbol{q},\sigma).
\end{equation}
Eq. (\ref{A6}) is just the usual behavior of a propagator near its
pole except for the factor $|N|^2$, and we can redefine a normalized
field to absorb this factor
\begin{equation}\label{A8}
(\Psi_0,\psi_l(0)\Psi_{\boldsymbol{q},\sigma})=(2\pi)^{-3/2}u_l(\boldsymbol{q},\sigma).
\end{equation}
This normalized field is called a renormalized field.  On the other
hand the mass $m$ in the pole also need not to correspond to a
elementary field, so we call $m$ the renormalized mass, which is
defined by the position of the pole.

Here we take a theory of scalar field for example, the free field is
$\Psi_B$, with mass $m_B$. By introducing a renormalized field
$\Phi\equiv Z^{-1/2}\Phi_B$ and mass $m^2\equiv m_B^2+\delta m^2$,
we can rewrite the Lagrangian density as
\begin{equation}\label{A9}
\begin{aligned}
&\mathscr{L}=\mathscr{L}_0+\mathscr{L}_1,
\\&\mathscr{L}_0=-\partial_{\mu}\Phi^{-}\partial^{\mu}\Phi^{+}-m^2\Phi^{-}\Phi^{+}
\\&\mathscr{L}_1=-(Z-1)[\partial_{\mu}\Phi^{-}\partial^{\mu}\Phi^{+}+m^2\Phi^{-}\Phi^{+}]\\&~~~~~~~+Z\delta m^2\Phi^{-}\Phi^{+}-V(\Phi).
\end{aligned}
\end{equation}
Now we can use OPI to calculate the full propagator
\begin{equation}\label{A10}
\begin{aligned}
\Delta'(q)=&[-2\sqrt{\boldsymbol{q}^2+m^2}(q^0-\sqrt{\boldsymbol{q}^2+m^2}+i\epsilon)]^{-1}\\&+[-2\sqrt{\boldsymbol{q}^2+m^2}(q^0-\sqrt{\boldsymbol{q}^2+m^2}+i\epsilon)]^{-1}\Pi(q)\\&\times[-2\sqrt{\boldsymbol{q}^2+m^2}(q^0-\sqrt{\boldsymbol{q}^2+m^2}+i\epsilon)]^{-1}+\cdots
\\=&[-2\sqrt{\boldsymbol{q}^2+m^2}(q^0-\sqrt{\boldsymbol{q}^2+m^2}+i\epsilon)-\Pi(q)]^{-1}.
\end{aligned}
\end{equation}
In calculating OPI graph $\Pi$, we encounter a vertices contribution
arising from $\mathscr{L}_1$, plus a loop term $\Pi_{\rm LOOP}$
\begin{equation}\label{A11}
\Pi(q)=-(Z-1)(q^2+m^2)+Z\delta m^2+\Pi_{\rm LOOP}(q).
\end{equation}
We can rewrite Eq. (A10) by Lorentz covariant terms
$\Delta^*(q)=(q^2+m^2-i\epsilon)$ and $\Pi^*(q^2)$
\begin{equation}\label{A12}
\Delta'(q)=[q^2+m^2-\Pi^*(q^2)-i\epsilon]^{-1},
\end{equation}
where
\begin{equation}\label{A13}
\begin{aligned}
\Pi^*(q^2)=&-(Z-1)(q^2+m^2)+Z\delta m^2+\Pi_{\rm
LOOP}(q)\\&-(\sqrt{\boldsymbol{q}^2+m^2}-q^0)^2.
\end{aligned}
\end{equation}
The non-covariant term $-(\sqrt{\boldsymbol{q}^2+m^2}-q^0)^2$ do not
contribute near the pole. In order to make the pole of the
propagator be at $q^2=-m^2$ and has a unit residue, we must have
\begin{equation}\label{A14}
\begin{aligned}
&\Pi^*(-m^2)=0,
\\&[\frac{d}{d q^2}\Pi^*(q^2)]_{q^2=-m^2}=0,
\end{aligned}
\end{equation}
which leads to
\begin{equation}\label{A15}
\begin{aligned}
&Z\delta m^2=-[\Pi_{\rm LOOP}(q)]_{q^0=\sqrt{\boldsymbol{q}^2+m^2}},
\\&Z=1+[\frac{d}{d q^0}\Pi_{\rm LOOP}(q)]_{q^0=\sqrt{\boldsymbol{q}^2+m^2}}.
\end{aligned}
\end{equation}
Similar analyses can also be applied to particles of arbitrary spin.

Then we see the $n=3$ case, one typical example is vertex
correction. We can define the vertex function $\Gamma^{\mu}$ of the
charged particle by
\begin{equation}\label{A16}
\begin{aligned}
\int& d^4x d^4y d^4z e^{-ip\cdot x}e^{-ik\cdot y}e^{+il\cdot
z}(\Psi_0,T\{J^{\mu}(x)\Psi^+_{n}(y){\bar
\Psi^+}_m(z)\}\Psi_0)\\&\rightarrow
-i(2\pi)^4qS'_{nn'}(k)\Gamma_{n'm'}^{\mu}(k,l)S'_{m'm}(l)\delta^4(p+k-l),
\end{aligned}
\end{equation}
in which $q$ is charge and $S'$ is the complete spinor propagator
\begin{equation}\label{A17}
\begin{aligned}
 &-i(2\pi)^4S'_{nm}(k)\delta^4(k-l)\\\leftarrow&\int d^4y
d^4z(\Psi_0,T\{\Psi^+_n(y){\bar \Psi^+}_m(z)\}\Psi_0)e^{-ik\cdot
y}e^{+il \cdot z}.
\end{aligned}
\end{equation}
We can see that $\Gamma^{\mu}$ is the sum of vertex graphs with one
incoming spinor line, one outgoing spinor line and one vector line
(with the external line coefficients stripped away), in the limit of
no interactions this becomes $\gamma^{\mu}$.

In conventional field theory, we can give a relation between
$\Gamma^{\mu}$ and $S'$ through Ward identity. Here we will show
that  a similar relation also exists in infinite statistics theory.
First we have
\begin{equation}\label{A18}
\begin{aligned}
\frac{\partial}{\partial x^{\mu}}T&\{J^{\mu}(x)\Psi^+_{n}(y){\bar
\Psi^+}_m(z)\}=T\{\partial_{\mu}J^{\mu}(x)\Psi^+_{n}(y){\bar
\Psi^+}_m(z)\}
\\&+\delta(x^0-y^0)T\{[J^{0}(x),\Psi^+_{n}(y)]{\bar \Psi^+}_m(z)\}\\&+\delta(x^0-z^0)T\{\Psi^+_{n}(y)[J^{0}(x),{\bar \Psi^+}_m(z)]\},
\end{aligned}
\end{equation}
where the delta functions arise from time-derivatives of step
functions. The first term of the right hand side vanishes by the
conservation condition $\partial_{\mu}J^{\mu}=0$. From Eqs.
(\ref{A16}) and (\ref{A17}) we can see that the real part in
$J^{\mu}$ that gives contributions is $-iq{\bar
\Psi^+}\Gamma^{\mu}\Psi^+$, so
\begin{equation}\label{A19}
\begin{aligned}
\left[J^0(\boldsymbol{x},t),\Psi^+_n(\boldsymbol{y},t)\right]=&-q\Psi^{+\dag}(\boldsymbol{x},t)\Psi^+(\boldsymbol{x},t)\Psi^+_n(\boldsymbol{y},t)\\&+q\Psi^+_n(\boldsymbol{y},t)\Psi^{+\dag}(\boldsymbol{x},t)\Psi^+(\boldsymbol{x},t),
\end{aligned}
\end{equation}
The first term will be annihilated by the vacuum state when
substituted into Eqs. (\ref{A18}) and (\ref{A16}), while the second
term becomes
$-q\Psi_n(\boldsymbol{y},t)\delta^3(\boldsymbol{x}-\boldsymbol{y})$
on the mass shell, so we can get
\begin{equation}\label{A20}
\begin{aligned}
&\left[J^0(\boldsymbol{x},t),\Psi^+_n(\boldsymbol{y},t)\right]\rightarrow-q\Psi_n^+(\boldsymbol{y},t)\delta^3(\boldsymbol{x}-\boldsymbol{y}),
\\&\left[J^0(\boldsymbol{x},t),{\bar \Psi^+}_n(\boldsymbol{y},t)\right]\rightarrow q{\bar \Psi^+}_n(\boldsymbol{y},t)\delta^3(\boldsymbol{x}-\boldsymbol{y}).
\end{aligned}
\end{equation}
Substituting  this into Eq. (\ref{A18}) gives
\begin{equation}\label{A21}
\begin{aligned}
\frac{\partial}{\partial x^{\mu}}T\{J^{\mu}(x)\Psi^+_{n}(y)&{\bar
\Psi^+}_m(z)\}\rightarrow-q\delta^4(x-y)T\{\Psi^+_{n}(y){\bar
\Psi^+}_m(z)\} \\&+q\delta^4(x-z)T\{\Psi^+_{n}(y){\bar
\Psi^+}_m(z)\}.
\end{aligned}
\end{equation}
From Fourier transform (\ref{A16}) we have Ward identity
\begin{equation}\label{A22}
(l-k)_{\mu}\Gamma^{\mu}(k,l)=i{S'}^{-1}(k)-i{S'}^{-1}(l).
\end{equation}
In the $l\rightarrow k$ limit, this gives
\begin{equation}\label{A23}
\Gamma^{\mu}(k,k)=-i\frac{\partial}{\partial k_{\mu}}{S'}^{-1}(k).
\end{equation}
As we discussed above, $S'$ can be written as a Lorentz covariant
form
\begin{equation}\label{A24}
{S'}^{-1}=i\partial {k\!\!\!\slash}+m-\Sigma^*({k\!\!\!\slash}),
\end{equation}
in which ${k\!\!\!\slash}=k^{\mu}\gamma_{\mu}$, then Eq. (\ref{A23})
becomes
\begin{equation}\label{A25}
\Gamma^{\mu}(k,k)=\gamma^{\mu}+i\frac{\partial}{\partial
k_{\mu}}\Sigma^*({k\!\!\!\slash}).
\end{equation}
The condition that the pole of $S'$ has a unit residue requires
$\frac{\Sigma^*({k\!\!\!\slash})}{\partial
{k\!\!\!\slash}}|_{{k\!\!\!\slash}=im}=0$, so on the mass shell we
have
\begin{equation}\label{A26}
\bar{u}'_k\Gamma^{\mu}(k,k)u_k=\bar{u}'_k\gamma^{\mu}u_k,
\end{equation}
in which
$[i\gamma_{\mu}k^{\mu}+m]u_k=[{i\gamma_{\mu}k^{\mu}+m}]u'_k=0$.

For the vector particle self-energy graph, there is a bit different.
In conventional field theory, we can introduce
\begin{equation}\label{A27}
M^{\mu\nu}(q)=\int d^4xd^4y e^{-iq\cdot x}e^{-iq'\cdot
y}(\Psi_0,T\{J^{\mu}(x)J^{\nu}(y)\}\Psi_0)
\end{equation}
to construct the complete propagator $\Delta'$
\begin{equation}\label{A28}
\begin{aligned}
\Delta'_{\mu\nu}(q)&=\Delta_{\mu\nu}+\Delta_{\mu\rho}(q)M^{\rho\sigma}(q)\Delta_{\sigma\nu}\\&=\Delta_{\mu\nu}+\Delta_{\mu\rho}(q)\Pi^{\rho\sigma}(q)\Delta'_{\sigma\nu}.
\end{aligned}
\end{equation}
From current conservation and local commutativity, we can get
$q^{\mu}M_{\mu\nu}(q)=0$, then from the specific form of
$\Delta_{\mu\nu}$, we have
$q^{\mu}\Delta'_{\mu\nu}(q)=q^{\mu}\Delta_{\mu\nu}(q)$, which leads
to $q_{\rho}\Pi^{\rho\sigma}(q)=0$. While in infinite statistics
theory, we can not solve the problem in the same way because of the
lake of locality. However, since the Lorentz covariant part in each
order of Feynman graph has the same form as that in conventional
theory and vanishes when contracting with $q$, we guess that for
infinite statistics OPI, the covariant part $\Pi^*(q)$ satisfies
$q_{\rho}\Pi^{*\rho\sigma}(q)=0$, then we can get the form of
$\Pi^{\mu\nu}(q)$ (see Eq. (\ref{31})).

\end{document}